\documentclass[aps,prd,superscriptaddress,twocolumn,nofootinbib,amsmath,amssymb,groupedaddress]{revtex4}
\usepackage{amsmath}
\usepackage{graphicx}
\usepackage{latexsym}
\usepackage{bm}
\usepackage{array}
\usepackage{dcolumn}
\usepackage{longtable}
\usepackage{tabularx}
\usepackage[usenames,dvipsnames]{color}
\newcommand{\be}{\begin{equation}}\newcommand{\ee}{\end{equation}}
\newcommand{\bea}{\begin{eqnarray}}\newcommand{\eea}{\end{eqnarray}}
\newcommand{\brr}{\begin{array}}\newcommand{\err}{\end{array}}
\newcommand{\bit}{\begin{itemize}}\newcommand{\eit}{\end{itemize}}
\newcommand{\ben}{\begin{enumerate}}\newcommand{\een}{\end{enumerate}}

\newcommand{\bbm}{\begin{bmatrix}}\newcommand{\ebm}{\end{bmatrix}}
\newcommand{\ba}{\begin{array}}
\newcommand{\ea}{\end{array}}

\newtheorem{mydef}{Definition}
\newtheorem{Lemma}{Lemma}
\newtheorem{theorem}{Theorem}
\newcommand{\bd}{\begin{mydef}} \newcommand{\ed}{\end{mydef}}
\newcommand{\bthe}{\begin{theorem}} \newcommand{\ethe}{\end{theorem}}
\newcommand{\ble}{\begin{Lemma}} \newcommand{\ele}{\end{Lemma}}

\def\Tr{\mathrm{Tr}}
\def\ph{\varphi}
\def\lan{\langle}
\def\lf{\left}

\def\non{\nonumber}\def\ran{\rangle}

\def\ri{\right}
\def\al{\alpha}\def\ga{\gamma}
\def\de{\delta}\def\De{\Delta}

\def\si{\sigma}\def\Si{\Sigma}

\def\1{{_{1}}}\def\2{{_{2}}}

\newcommand{\ide}{1\hspace{-1mm}{\rm I}}

\def\noHe0{:\;\!\!\;\!\!:H_e(0):\;\!\!\;\!\!:}
\def\noHm0{:\;\!\!\;\!\!:H_\mu(0):\;\!\!\;\!\!:}

\def\lan{\langle}
\def\lf{\left}

\def\non{\nonumber}
\def\ran{\rangle}

\def\ri{\right}

\def\al{\alpha}\def\ga{\gamma}
\def\de{\delta}\def\De{\Delta}

\def\si{\sigma}\def\Si{\Sigma}

\def\1{{_{1}}}\def\2{{_{2}}}
\def\I{{_{\rm{I}}}}\def\II{{_{\rm{II}}}}

\definecolor{darkred}{rgb}{.8,0,0}

\definecolor{darkblue}{rgb}{0,0,.7}

\begin{document}

\title{Bekenstein bound from the Pauli principle\footnote{To our friend, colleague and inspirer Martin Scholtz (1984 - 2019).}}
\author{G.~Acquaviva}
\email{gioacqua@utf.troja.mff.cuni.cz}

\affiliation{Faculty  of  Mathematics  and  Physics, Charles  University, V  Hole\v{s}ovi\v{c}k\'{a}ch  2, 18000  Praha  8,  Czech  Republic.}

\author{A.~Iorio}
\email{iorio@ipnp.troja.mff.cuni.cz}

\affiliation{Faculty  of  Mathematics  and  Physics, Charles  University, V  Hole\v{s}ovi\v{c}k\'{a}ch  2, 18000  Praha  8,  Czech  Republic.}

\author{L.~Smaldone}
\email{smaldone@ipnp.mff.cuni.cz}

\affiliation{Faculty  of  Mathematics  and  Physics, Charles  University, V  Hole\v{s}ovi\v{c}k\'{a}ch  2, 18000  Praha  8,  Czech  Republic.}

\begin{abstract}
Assuming that the degrees of freedom of a black hole are finite in number and of fermionic nature, we naturally obtain, within a second-quantized toy model of the evaporation, that the Bekenstein bound is a consequence of the Pauli exclusion principle for these fundamental degrees of freedom. We show that entanglement, Bekenstein and thermodynamic entropies of the black hole all stem from the same approach, based on the entropy operator whose structure is the one typical of Takahashi and Umezawa's Thermofield Dynamics.
We then evaluate the von Neumann black hole--environment entropy and noticeably obtain a Page-like evolution. We finally show that this is a consequence of a duality between our model and a quantum dissipative-like fermionic system.
\end{abstract}

\vspace{-1mm}

\maketitle
\section{Introduction}

This paper moves from the results of previous research \cite{Acquaviva:2017xqi}, but reversing the point of view. There, Bekenstein's argument that a black hole (BH) reaches the maximal entropy at disposal of a physical system (i.e., that it saturates the Bekenstein bound \cite{BB}), leads to two main proposals: i) the degrees of freedom (dof) responsible for the BH entropy have to take into account both matter and spacetime and hence must be of a new, more fundamental nature than the dof we know (with Feynman \cite{Feynman}, here we call such dof ``$X$ons'', see also \cite{Bao:2017rnv} and \cite{bekensteinSciAm}); ii) the Hilbert space $\mathcal{H}$ of the $X$ons of a given BH is necessarily finite dimensional
\be \label{dimhe}
\mathrm{dim} \mathcal{H}=e^{{\cal S}_{BH}} \, ,
\ee
with ${\cal S}_{BH}$ the Bekenstein entropy. With these, in \cite{Acquaviva:2017xqi} it was shown that the (average) loss of information is an unavoidable consequence of the non-vanishing relic entanglement between the evaporated matter and the spacetime.

In search of a unifying view of the various types of entropies involved in the BH evaporation (i.e., Bekenstein, thermodynamical, and entanglement entropies, see, e.g., \cite{HarlowReview}), here we reverse that logic. Namely, we start off by supposing that in a BH only free $X$ons exist (hence there can only be one kind of entropy at that level), and we suppose that they are \textit{finite} in number and \textit{fermionic} in nature. This amounts to have a finite dimensional $\mathcal{H}$. With these assumptions, here we show that the evaporation is a dynamical mechanism producing a maximal entanglement entropy, equal to the initial entropy of the BH.

This is an instance of the Bekenstein bound, obtained here with arguments that do not assume pre-existing geometrical (spatiotemporal) concepts. In fact, for a full identification with the standard formulae (see \cite{Casini_2008}, where the bound is rigorously defined in quantum field theory (QFT) and, e.g., the review \cite{Bousso:2002ju})  one needs to associate geometrical concepts to the $X$ons. For instance, one could make each dof correspond to one elementary Planck cell.
Nonetheless, in our picture we do not need the exact expression of the bound. What is crucial is that the $X$ons are taken to be finite in number and fermionic, otherwise the entanglement entropy would just indefinitely grow without reaching a maximal value. It is suggestive, though, that taking on board the geometric picture of $X$ons as quanta of area (Planck cells), the horizon of the BH is of nonzero size as an effect of a Pauli exclusion principle. Before entering the details of what just discussed, let us now briefly put our work into the context of current literature.

Bekenstein entropy \cite{Bekenstein:1972tm,Bekenstein:1973ur} is traditionally regarded as a measure of our ignorance about the dof which formed the BH \cite{Bekenstein:1973ur, Hawking:1976de,Zurek:1982zz,Page:1983ug} and as a consequence of the \emph{no-hair theorem} \cite{nohair}. However, other interpretations have been proposed in literature, as in \emph{Loop Quantum Gravity} (LQG), where BH entropy is a counting of microstates corresponding to a given macroscopic horizon area $\mathcal{A}$ \cite{Ashtekar:1997yu, Rovelli:2004tv}. Along these lines, Bekenstein proposed a universal upper bound for the entropy of any physical system contained in a finite region \cite{BB}, which is saturated by BHs. This implies \cite{Diaz} that the entropy of every system in a finite region is bounded from above by the Bekenstein entropy of a BH, whose horizon coincides with the area of the boundary of that region (see \cite{Casini_2008}, and also \cite{Bousso:2002ju}).

Using the approach of QFT in curved spacetimes, Hawking discovered the black body spectrum of BH radiation~\cite{Hawking}. In the meantime, Umezawa and Takahashi developed their \emph{Thermofield dynamics} (TFD) \cite{UmeTak} (see also Ref. \cite{Umezawa}), that immediately appeared to be a fruitful tool to describe BH evaporation \cite{Israel}. In \cite{IorLamVit}, with the help of an \textit{entropy operator}, whose structure is natural in TFD, the BH-radiation entropy is viewed as an entanglement entropy of radiation modes with their ``TFD-double'' (the modes beyond the horizon).

Although the relation between QFT in curved spacetime and TFD was studied already in Refs. \cite{MarSodVit, Laciana:1994wd}, the renewed interest comes in connection with the AdS/CFT correspondence \cite{Maldacena:2001kr}, where in a \textit{two-sided} Anti-de Sitter (AdS) BH, the specular asymptotic region is mapped into \textit{two copies} of a conformal field theory (CFT). The thermal nature of the BH is then naturally seen through TFD. Extensions to incorporate dissipative effects are in the recent \cite{BottaCantcheff:2017kys, Dias:2019ezx}.

Since a BH, initially described as a pure state, could end up in a mixed state (this is actually the view of \cite{Acquaviva:2017xqi}), questions arise on the unitary evolution, as first noticed by Hawking \cite{Haw1976} and then extensively discussed, from different points of view, see e.g. Refs. \cite{Page:1979tc, Giddings:1992ff, Page:1993wv, tHooft:1995qox,Hawking:2005kf, Hayden:2007cs, Mathur:2009hf, Page:2013dx, Polchinski:2016hrw,AlAlvi:2019vzd, Piroli:2020dlx}. In particular, in Refs. \cite{Page:1993wv, Page:1993df} Page studied the bipartite system BH-radiation, in a random (Haar distributed) pure state, computing the radiation entanglement entropy as function of the associated thermodynamical entropy. He found a symmetric curve (\emph{Page curve}) which goes back to zero when the BH is completely evaporated. In Ref. \cite{Page:2013dx} he postulated that entanglement entropy, as function of time, follows the minimum between Bekenstein and radiation thermodynamic entropy (\emph{Conjectured Anorexic Triangular Hypothesis}). Recently \cite{Almheiri:2019hni}, Page curve was also derived from holographic computations \cite{Ryu:2006bv}.

As said, in this paper we reverse the line of reasoning of Ref. \cite{Acquaviva:2017xqi} and present a simple, purely quantum toy-model of the dynamics of BH evaporation, focusing on the fundamental dof. In Section \ref{BasicAssumption} the basic assumptions are the finiteness of slots (quantum levels) available for the system, and the fermionic nature of such dof. The finiteness of the Hilbert space of states follows from the Pauli exclusion principle. In Section \ref{BHevaporation} we compute the von Neumann entropy of the subsystems during their evolution. This is remarkably given by the expectation value of the TFD entropy operator \cite{IorLamVit} and it has the same qualitative behavior of Page curve: it starts from zero and ends in zero, while its maximum is reached at half of the evaporation process. That maximum is identified here with the Bekenstein entropy of the BH at the beginning of the evaporation. We can therefore argue that Bekenstein bound itself descends from the Pauli principle. In Section \ref{QuantumDissipation} we explain the relation with TFD by mapping our model to an equivalent description as a dissipative-like system. The last Section is left to our conclusions, while in the Appendix we show the connection between TFD and von Neumann entropies in the present context.

Throughout the paper we adopt units in which $c=\hbar=1$.

\section{Basic assumptions and model of BH evaporation}\label{BasicAssumption}
We assume that the fundamental dof are fermionic (BH and models based on fermions are available in literature, see, e.g., the \emph{SYK} model \cite{Sachdev:2015efa, Maldacena:2016hyu}). As a consequence, each quantum level can be filled by no more than one fermion. This assures that the Hilbert space $\mathcal{H}$ of physical states with a finite number of levels is finite dimensional. In fact, if the fundamental modes were bosons, the requirement that the number of slots available were finite would not have been sufficient to guarantee the finiteness of $\mathrm{dim} \,\mathcal{H}$.
Let us recall now that, in the picture of \cite{Acquaviva:2017xqi}, it is only at energy scales below those of quantum gravity (e.g., at the energy scales of ordinary matter) that the field modes are distinguishable from those ``making'' the spacetime, hence we can write
\be
  \mathcal{H}_F \otimes \mathcal{H}_G \ \subseteq \ \mathcal{H} \, .
\ee
Here $F$ and $G$ stand for ``fields'' and ``geometry'', respectively. In other words, at low energy, the $F$-modes will form quantum fields excitations, that is, the \textit{quasi-particles} (from the $X$ons point of view) immersed into the spacetime formed by the $G$-modes.

Now, say $N$ is the total number of quantum levels (slots) available to the BH. The evaporation consists of the following, steady process: $N \to (N -1) \to (N - 2) \to \cdots$. That is, the number of \textit{free} $X$ons steadily decreases, in favor of the $X$ons that, having evaporated, are arranged into quasi-particles and the spacetime they live in. One might think of a counter that only sees free $X$ons, hence keeps clicking in one direction as the BH evaporates, till its complete stop.

In this picture: i) there is no pre-existing \textit{time}, because the natural evolution parameter is the average number of free $X$ons; and ii) there is no pre-existing \textit{space} to define the regions inside and outside the BH, because a distinction of the total system into two systems, say \emph{environment} ($\mathrm{I}$) and \emph{BH} ($\mathrm{II}$), naturally emerges in the way just depicted.
With this in mind, in what follows we shall nonetheless refer to the system $\mathrm{I}$ as \textit{outside}, and to system $\mathrm{II}$ as \textit{inside}. It is a worthy remark that other authors do use the geometric notion of \emph{exterior} and \emph{interior} of BH, even at fundamental level \cite{Page:1979tc}. Even though this can be justified, see, e.g., \cite{Bao:2017rnv}, and permits to produce meaningful models, see, e.g., \cite{Piroli:2020dlx}, our approach does not require it.
The Hilbert space of \textit{physical states} is then built as a subspace of a larger tensor product (\textit{kinematical}) Hilbert space
\be
\mathcal{H} \ \subseteq \ \mathcal{H}_\I \otimes \mathcal{H}_\II \,.
\ee
We now assume that such a Hilbert space can be constructed with the methods of second quantization. This provides a language contiguous to the language of QFT, which should be recovered in some limit. Therefore, BH and environment modes will be described by two sets of creation and annihilation operators, which satisfy the usual canonical anticommutation relations
\be
\lf\{\chi_{\tau n},\chi^\dag_{\tau'n'}\ri\} \ = \ \de_{\tau,\tau'} \, \de_{n n^{'}} \, ,
\ee
with $n,n'= 1, \dots, N$, $\tau \ = \ \mathrm{I} ,  \mathrm{II}$, and all other anticommutators equal to zero. Then, we introduce the simplified notation
\be\label{opfac}
\chi_{\I\, n} \ = \ a_n \otimes \ide_\II \equiv a_n \,, \qquad \chi_{\II \, n} \ =  \ide_\I \otimes b_n \equiv b_n\, .
\ee
We suppose that the initial state is
\be \label{instate}
|0,N\ran \ \equiv \ |0,0,\ldots,0\ran_{\I} \otimes |1,1,\ldots,1\ran_{\II} \, ,
\ee
where both kets, $\mathrm{I}$ and $\mathrm{II}$, have $N$ entries and
\be
|1,1,\ldots,1\ran_{\II} \ = \ b^{\dag}_{1} b^{\dag}_{2} \ldots b^{\dag}_{N} \, |0,0,\ldots,0\ran_{\II} \, .
\ee
The state in Eq.\eqref{instate} represents the BH at the beginning of the evaporation process, with all the slots occupied by free $X$ons. Although the $X$ons, during the evaporation, are progressively arranged into less fundamental structures (and hence no longer are the dof to be used for the emergent description) we keep our focus on them. For us this ``transmutation'' only helps identifying what to call ``inside'' and what  ``outside'', so that evaporation is the process that moves the $X$ons from $\mathrm{II}$ to $\mathrm{I}$.
In this way, the final state (for which there are no free $X$ons left, as they all recombined to form fields and spacetime), has the form
\be \label{finstate}
|N,0\ran \ \equiv \ |1,1,\ldots,1\ran_{\I} \otimes  |0,0,\ldots,0\ran_{\II} \, ,
\ee
where
\be
|1,1,\ldots,1\ran_{\I} \ = \ a^{\dag}_{1} a^{\dag}_{2} \ldots a^{\dag}_{N} \, |0,0,\ldots,0\ran_{\I} \, .
\ee

In order to construct a state of the system, compatible with the previous assumptions, let us consider the evolved operators as
\bea \label{rot1}
c_n(\si) & = & e^{i \psi_n} \, \lf(b_n \, \cos \si \, + \, a_n \, e^{-i \ph_n} \sin \si \ri)\, , \\[2mm]
d_n(\si) & = & e^{i \psi_n} \, \lf(e^{-i \ph_n} a_n \, \cos \si \, - \, b_n \, \sin \si\ri) \,,  \label{rot2}
\eea
where on $\sigma$ we shall comment soon. Eqs.\eqref{rot1} and \eqref{rot2} define a canonical transformation
\be
\lf\{c_n(\si) \, , \, c^\dag_m(\si) \ri\} \ = \ \lf\{d_n(\si) \, , \, d^\dag_m(\si) \ri\} \ = \ \de_{nm} \, .
\ee
We thus get the evolved of the initial state \eqref{instate} as
\bea \label{bhstate}
&& |\Psi(\si)\ran \ \equiv \ \prod^N_{n=1} \, c^\dag_n (\si) \, |0\ran_{\I} \otimes |0\ran_{\II} \\[2mm] \non
&& = \ \prod^N_{n=1} \,e^{-i \psi_n} \lf( b^\dag_{n} \, \cos \si \, + \, a^\dag_{n} \, e^{i \ph_n} \, \sin \si \ri)|0\ran_{\I} \otimes |0\ran_{\II} \, .
\eea
Strictly speaking, $\si$ should be regarded as a discrete parameter, counting the free $X$ons that leave the BH, according to the picture above described (see also the discussion in the next Section). Nonetheless, in order to simplify computations, and with no real loss of generality, we use the continuous approximation.  Given our initial state (Eq.\eqref{instate}) and final state (Eq.\eqref{finstate}), $\si$ can be seen as an interpolating parameter, describing the evolution of the system, from $\si = 0$, corresponding to the beginning of the evaporation, till $\si = \pi/2$, corresponding to complete evaporation.

Let us also notice that the linear canonical transformation defined in Eqs.\eqref{rot1},\eqref{rot2} is very general, given the requests. In fact, if we mix creation and annihilation operators, $c_n (\si) \sim (a_n + b^\dag_n)$, one cannot interpolate Eqs.\eqref{instate} and \eqref{finstate}. Furthermore, the choice of phases introduced does not affect any of the results presented. This is a consequence of the fact that we are working with \emph{two} types of modes (BH and environment). If we had more than two, we would have to deal with one or more physical phases, as is well known in quark and neutrino physics \cite{maskawa}. We can thus safely set $\ph_n \ = \ 0 \ = \psi_n$.

With our choice of parameters, the state \eqref{bhstate}, can also be written as
\be \label{alexp}
|\Psi(\si)\ran \ = \ \prod^N_{i=1} \, \sum_{n_i=0,1} \, C_i(\si) \, \lf(a^{\dag}_i\ri)^{n_i} \, \lf(b^{\dag}_i \ri)^{1-n_i}  \, |0\ran_{\I} \otimes |0\ran_{\II} \, ,
\ee
with $C_i= (\sin \si)^{n_i} \, (\cos \si)^{1-n_i}$. This form would suggest the following generalization
\be
|\Phi(\si)\ran \ = \ \prod^N_{i=1} \, \sum_{n_i, m_i=0,1} \, D_i(\si) \, \lf(a^{\dag}_i\ri)^{n_i} \, \lf(b^{\dag}_i \ri)^{m_i}  \, |0\ran_{\I} \otimes |0\ran_{\II} \, ,
\ee
with $D_i= (\sin \si)^{n_i} \, (\cos \si)^{m_i}$. However, we easily compute
\bea
|\Phi(0)\ran & = & |0_1,\ldots,0_N\ran_{\II} \otimes |1_1,\ldots,1_N\ran_{\II}\non \\[2mm]
& + & |0_1,\ldots,0_N\ran_{\II} \otimes |0_1,\ldots,0_N\ran_{\II} \, .
\eea
which is incompatible with our boundary condition \eqref{instate}. In order to enforce the latter, we need to impose the constraint $m_i=1-n_i$.

\section{Entropy operators, Page curve and the Bekenstein bound} \label{BHevaporation}

The Hilbert space of physical states has dimension
\be \label{hdim}
\Si \ \equiv \ {\rm dim} \, \mathcal{H} \ = \ 2^N  \, .
\ee
The state defined in Eq.\eqref{bhstate} is an entangled state. This is due to the fact that $c^\dag_n(\si)$ cannot be factorized as $a_n$ and $b_n$ in Eq.\eqref{opfac}, i.e. it cannot be written as $c^\dag_n \ = \ A_\I \otimes B_\II$, where $A_\I$ ($B_\II$) acts only on $\mathcal{H}_\I$ ($\mathcal{H}_\II$).

To quantify such entanglement we define the entropy operator for environment modes as in TFD \cite{UmeTak,Umezawa,IorLamVit}
\be
S_{\I}(\si) \ = \ -\sum^N_{n=1} \, \lf(a^\dag_n \, a_n \, \ln \sin^2 \si + a_n \, a^\dag_n \, \ln \cos^2 \si\ri) \, ,
\ee
We also define the entropy operator for BH modes, in a rather unconventional way
\be \label{s2}
S_{\II}(\si) \ = \ -\sum^N_{n=1} \, \lf(b^\dag_n \, b_n \, \ln \cos^2 \si + b_n \, b^\dag_n \, \ln \sin^2 \si\ri) \, .
\ee
The reason for such unconventional definition will be clear in the next Section. For the moment, notice that we have two different operators, for $\mathrm{I}$ and for $\mathrm{II}$, but we see that, since
\bea  \label{numav}
\lan a^\dag_n \, a_n \ran_\si \ & = & \ \sin^2 \si = 1- \lan b^\dag_n \, b_n\ran_\si \, ,
\eea
then
\bea
&& \mathcal{S}_\I (\si) \ \equiv \  \lan S_{\I}(\si) \ran_\si \non \\[2mm]
&& = \ -N \lf(\sin^2 \si \, \ln \sin^2 \si + \cos^2\si \, \ln \cos^2 \si\ri) \non \\[2mm]
&& = \ \lan S_{\II}(\si) \ran_\si \ \equiv \ \mathcal{S}_\II (\si)  \, , \label{s12}
\eea
where $\lan \ldots \ran_\si \equiv \lan \Psi(\si)|\ldots |\Psi(\si)\ran$.
Therefore the averages of the operators coincide, as it must be for a bipartite system. This entropy is the entanglement entropy between environment and BH, when the system evolves. Remarkably, it has a behavior in many respect similar to that of the Page curve \cite{Page:1993wv}, as shown in Fig. 1.
\begin{figure}[htb]
		\includegraphics[width=0.45\textwidth]{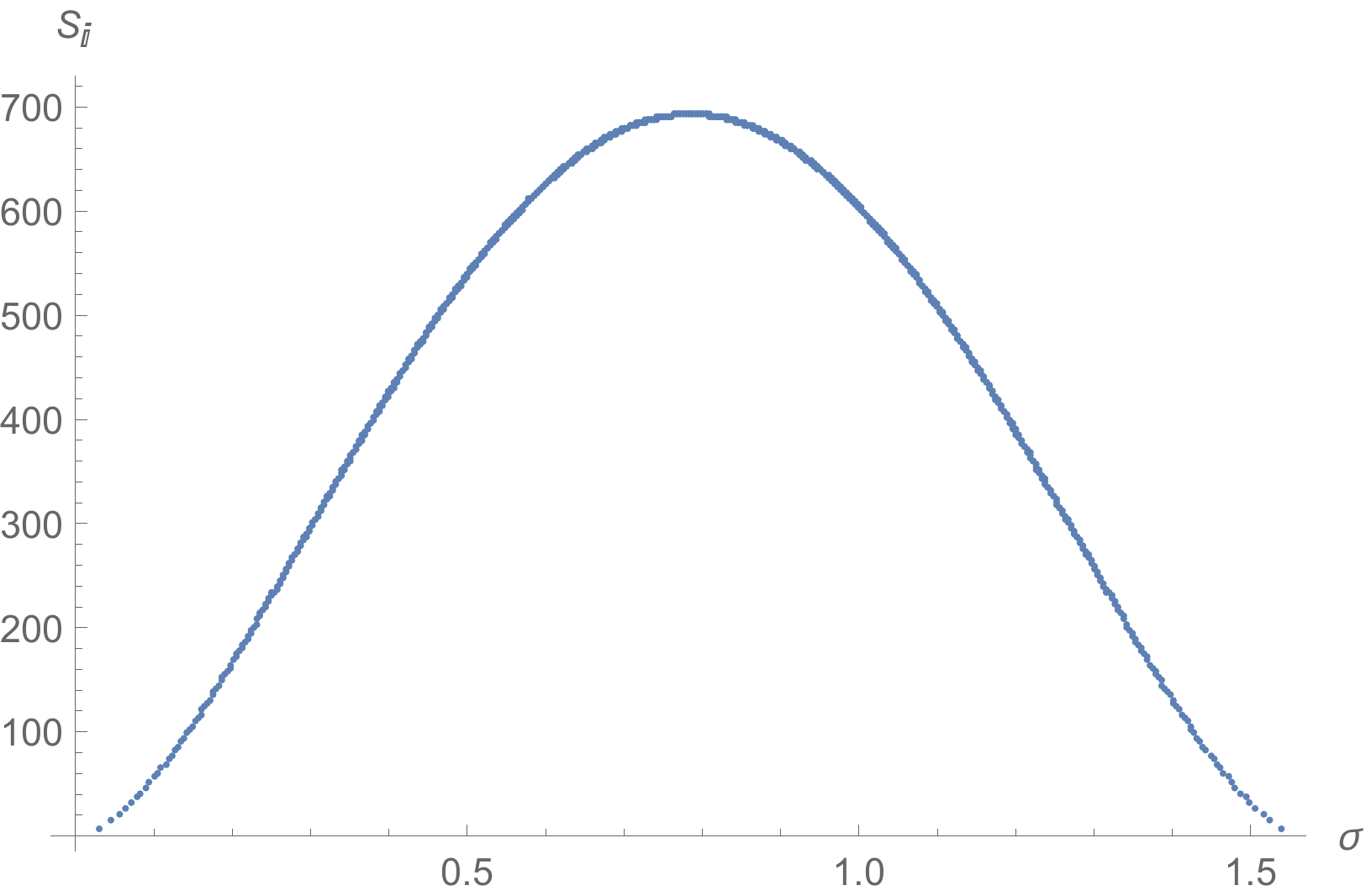}
	\label{fig:fentropy}
	\caption{The von Neumann entropy as a function of $\si$, in the case $N=1000$.}
\end{figure}

The maximum value is
\be \label{smax}
\mathcal{S}_{max} \ = \ N \, \ln 2 \ = \ \ln \Si \, ,
\ee
so that
\be \label{dimh}
\Si \ = \  e^{\mathcal{S}_{max}} \, .
\ee
As we see here through Eq. \eqref{dimh} (that is the analogue of Eq.\eqref{dimhe}), in our model $\mathrm{dim} \,\mathcal{H} $ is related to the maximal entanglement (von Neumann) entropy of the environment with the BH (and, of course, \textit{viceversa}). This happens exactly when the modes have \textit{half} probability to be inside and \textit{half} probability to be outside the BH\footnote{Recall that we have an intrinsic, non-geometric notion of the partition into inside/outside.}, and then a large amount of bits are necessary to describe the system. Thus, the system has an intrinsic way to know how big is the physical Hilbert space, hence to know how big is the BH at the beginning of the evaporation: when the maximal entanglement is reached, that value of the entropy, $\mathcal{S}_{max}$, tells how big was the original BH. Hence $\mathcal{S}_{max}$ must be some function of ${\cal M}_0, {\cal Q}_0, {\cal J}_0$,  i.e., the initial mass, charge or angular momentum of the BH.

This maximal entropy bound is obtained here as a mere consequence of the finiteness of the fermionic fundamental dof, hence of a Pauli principle. No geometric notions (distance, area, Planck length, etc) are employed. When such notions are eventually introduced, this bound \textit{must} correspond to the Bekenstein bound. In other words, the necessary \textit{dynamical map} connecting the $X$ons to fields and geometry will be introduced in such a way that this {\it fundamental, non-geometrical} bound becomes the {\it emergent, geometrical} Bekenstein bound. A brief discussion on the dynamical map is offered in the last section.

Therefore, for a full identification of ${\cal S}_{max}$ with ${\cal S}_{BH}$ we need more than what we have here. In particular, we need the concept of area, that somehow is what has been evoked in LQG \cite{Ashtekar:1997yu, Rovelli:2004tv} when in \eqref{smax} one identifies
\be \label{immirziLQG}
N \equiv \frac{ \cal A}{4 \pi \ga l_P^2 \sqrt{3}} \,,
\ee
where $\ga$ is the \emph{Immirzi parameter}.  We shall comment more on this later.

We want now to bring into the picture the two missing pieces: how the entropy of the BH, that should always decrease, and the entropy of the environment, that should always increase (hence, can be related to a standard thermodynamical entropy), actually evolve in our model. To this end, let us introduce the following number operators%
\be
\hat{N}_{\I} \ = \ \sum^N_{n=1} \, a^\dag_n \, a_n \, , \qquad \hat{N}_{\II} \ = \ \sum^N_{n=1} \, b^\dag_n \, b_n  \, ,
\ee
that count the number of modes of the radiation and the number of modes of the BH, respectively. Although it should be clear from the above, it is nonetheless important to stress now again that, in our formalism, the full kinematical Hilbert spaces associated to both sides have fixed dimension (${\rm dim} \, \mathcal{H}_\I = {\rm dim} \, \mathcal{H}_\II = 2^N$), while only a subspace $\mathcal{H} \subseteq  \mathcal{H}_\I \otimes  \mathcal{H}_\II$ such that ${\rm dim} \, \mathcal{H}= 2^N$ is the one of physical states. Note that $\mathcal{H}$ cannot be factorized and this is the origin of BH/environment entanglement.

Nonetheless, one could think that the physical Hilbert spaces of the two subsystems have to take into account only the number of modes truly occupied, at any given stage of the evaporation. Hence, the actual dimensions would be $2^{N_\I (\si)}$, and $2^{N_{\II} (\si)}$, where one easily finds that
\be
N_\I (\si) \ \equiv \ \lan  \hat{N}_{\I} \ran_\si = N \sin^2 \si \;, \label{n1}
\ee
and
\bea
N_\II (\si) & \equiv & \lan  \hat{N}_{\II} \ran_\si \non \\[2mm]
& = & N- N_\I(\si) \ = \ N \, \cos^2 \si \, . \label{n2}
\eea
Recall that $\sigma$ is, in fact, a discrete parameter, essentially counting the diminishing number of free $X$ons (as said earlier, and shown in greater detail later).

In other words, when we take this view, the partition into $\mathrm{I}$ and $\mathrm{II}$ becomes in all respects similar to the one of Page \cite{Page:1993wv}, that is
\be
2^N = 2^{N_{\II}(\si)} \times 2^{N_\I(\si)} \equiv n \times m \;,
\ee
with $n = 2^N, 2^{N - 1}, \dots, 1$, and $m = 1, \dots, 2^{N - 1}, 2^N$, while $\si$ runs in discrete steps in the interval $[0, \pi/2]$. Number fluctuations, which makes it necessary to invoke the entire Hilbert space $\mathcal{H}$ at each stage, represent a measure of entanglement of these modes, as we shall see below.
It is then natural to define the Bekenstein entropy as
\be \label{esbh}
{\cal S}_{BH} \equiv \ln n = N \ln 2 \cos^2 \sigma \,,
\ee
and the environment entropy\footnote{We could also call it \emph{thermodynamical entropy}, in comparison with the nomenclature of Ref. \cite{Page:1993wv}.} as
\be \label{eenv}
{\cal S}_{env} \equiv \ln m = N \ln 2 \sin^2 \sigma \,.
\ee

The plots of the three entropies, ${\cal S}_{I}$, ${\cal S}_{BH}$, ${\cal S}_{env}$ are shown in Fig. 2, and must be compared with similar results of Ref. \cite{Page:2013dx}. There are, though, two main differences worth stressing. First, we have a common single origin behind all involved entropies, as explained. Second, since the overall system here is based on the most fundamental entities, the curve for ${\cal S}_{I}$ cannot be always \textit{below} the other two, as happens in \cite{Page:2013dx}, but its maximum ${\cal S}_{max}$ must reach the starting point of ${\cal S}_{BH}$ (and the ending point of ${\cal S}_{env}$). In our case, the inequality
\be \label{ti}
\mathcal{S}_\I \ \leq \ \mathcal{S}_{BH} \, + \, \mathcal{S}_{env} \ = \  \mathcal{S}_{max}  \, ,
\ee
is always satisfied.
\begin{figure}[htbp]
	\centering
		\includegraphics[width=0.50\textwidth]{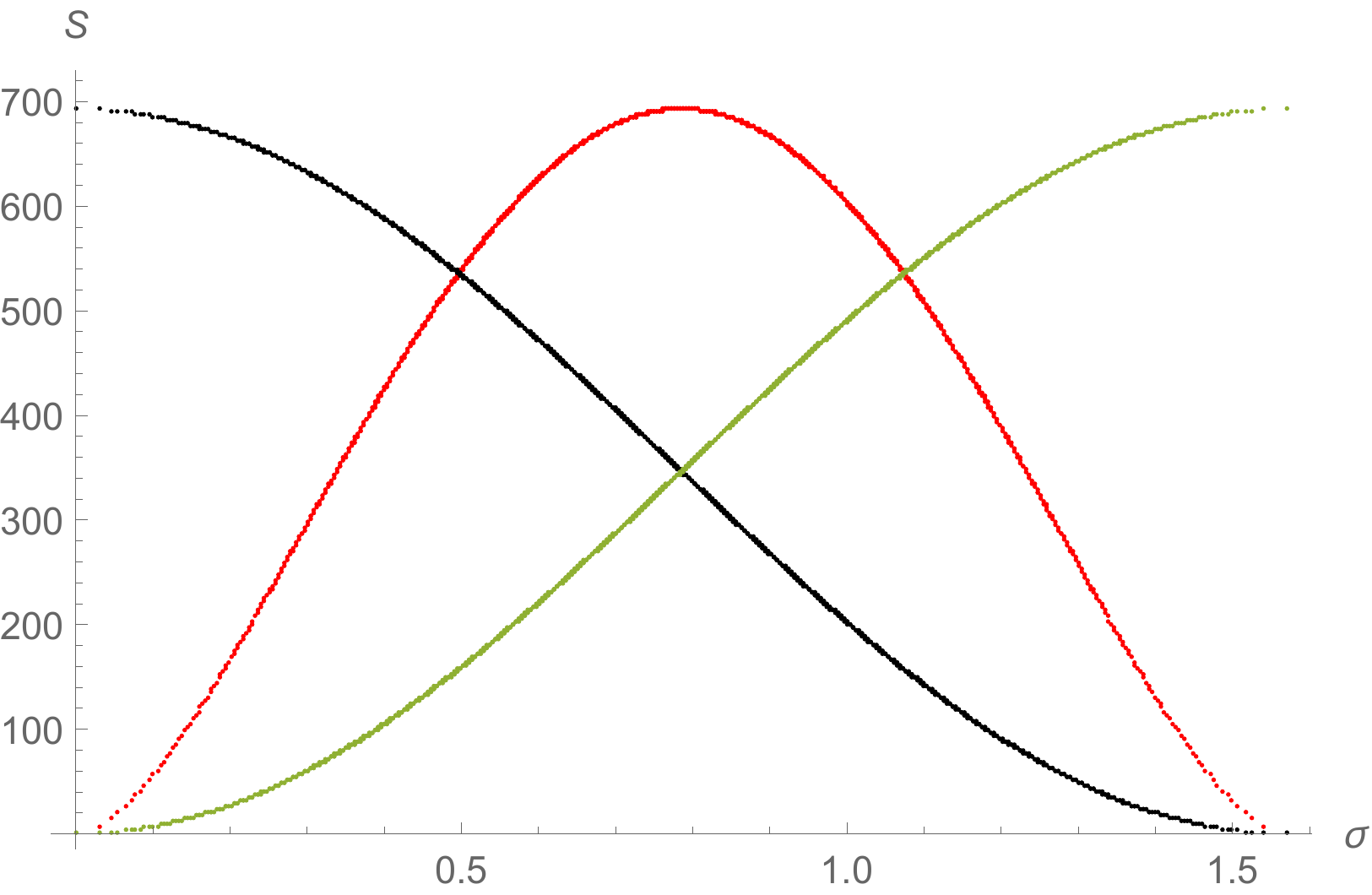}
	\label{fig:3entropies}
	\caption{Here we plot: $\mathcal{S}_{BH}$ in black, monotonically decreasing; $\mathcal{S}_{env}$ in green, monotonically increasing; $\mathcal{S}_{\I}$ in red, with a Page-like behavior. Note that the maximal value of $\mathcal{S}_{\I}$, $S_{max}$, coincides with the initial BH entropy, as well as with the final environment entropy, as inferred in the text. The plots are done for $N=1000$.}
\end{figure}
Note also that the dynamics of our system is unitary, because we keep our focus on the evaporated $X$ons, and not on the emerging structures, as was done in \cite{Acquaviva:2017xqi}. Hence we are not in the position here to spot the relic entanglement between fields and spacetime, that would make the curves for ${\cal S}_{BH}$ and ${\cal S}_{I}$ end at a nonzero value. The latter is precisely the source of the information loss in the quasi-particle picture of \cite{Acquaviva:2017xqi}. Whether or not this formally unitary evolution is physically tenable at the emergent level, and the impact of this on the validity of the Stone-von Neumann uniqueness theorem \cite{Umezawa, ineq,qSvN} in a quantum system with a finite-dimensional Hilbert space, is under scrutiny in ongoing research \cite{AIS_SvN}. Recently, the relation between unitarity and the existence of a maximal entropy has been also investigated in Ref. \cite{Dvali:2020wqi}. 

It is worth noticing that the total entropy $\mathcal{S}_{BH}+\mathcal{S}_{env}$ in Eq.\eqref{ti} is a constant along the evolution parameter $\sigma$: it is tempting to recognize this as a \emph{generalized second law} (GSL), saturated in this case, due to the fact that we are probing the fundamental, non-coarse-grained level of the Xons.  However we stress again here that $\sigma$ cannot be directly related to time evolution yet and hence a comparison with a  GSL can be premature.  Instead, one could expect that a proper GSL should arise from the coarse-grained description given by a dynamical map, something that was already hinted to in \cite{Acquaviva:2017xqi} due to the presence of the final residual entropy.

It is perhaps worthwhile to stress that Eq.\eqref{s12} represents exactly a von Neumann entropy. We can write the density matrix $\rho(\si) \ = \ |\Psi(\si)\ran \lan \Psi(\si)|$ and evaluate the reduced density matrices $\rho_{\I} = \Tr_{\II} \rho$ and $\rho_{\II} = \Tr_{\I} \rho$ \cite{NieChu}. We can then easily check that (see Appendix \ref{vntfd})
\bea
\mathcal{S}_\I (\si) & = & -\Tr_{\I} \lf( \rho_{\I}(\si) \ln \rho_{\I}(\si)\ri) \non \\[2mm]
& = & - N \lf(\cos^2 \si \, \ln \cos^2 \si+\sin^2 \si \, \ln \sin^2 \si\ri) \non \\[2mm]
& = & -\Tr_{\II} \lf( \rho_{\II}(\si) \ln \rho_{\II}(\si)\ri) \ = \ \mathcal{S}_\II (\si) \, . \label{svn}
\eea

Let us now consider some simple cases
\begin{itemize}
\item
If $N=1$ we have
\bea \label{1psi}
|\Psi(\si)\ran  & = &  \cos \si |0\ran_{\I} \otimes |1\ran_{\II} +  \sin \si |1\ran_{\I} \otimes |0\ran_{\II} \, .
\eea
In general, this is an entangled state, whose maximal entanglement is reached for $\si=\pi/4$:
\bea
|\Psi\lf(\pi /4\ri)\ran  =  \frac{1}{\sqrt{2}}\lf(|0\ran_{\I} \otimes |1\ran_{\II} +  \, |1\ran_{\I} \otimes |0\ran_{\II}\ri) \, .
\eea
\item
For $N=2$ we have
\bea
&& |\Psi(\si)\ran \ = \ \cos^2 \si \, |0_1 \, 0_2\ran_\I \otimes |1_1 \, 1_2\ran_\II  \non  \\[2mm]
&& + \  \sin^2 \si \,|1_1 \, 1_2\ran_\I \otimes |0_1 \, 0_2\ran_\II  \non  \\[2mm]
&&+  \, \cos \si \, \sin \si \,  |0_1 \, 1_2\ran_\I \otimes |1_1 \, 0_2\ran_\II \non \\[2mm]
&& +  \,   \cos \si \, \sin \si \,  |1_1 \, 0_2\ran_\I \otimes |0_1 \, 1_2\ran_\II \, . \label{2psi}
\eea
It is then clear that the mean number \eqref{numav} represents the \emph{probability of the $n$-th mode to ``leave the BH phase'' (to go from $\mathrm{II}$ to $\mathrm{I}$)}.
\end{itemize}

As mentioned earlier, $\si$ for us is a continuous approximation of a discrete parameter, that counts the $X$ons transmuting from being free (in the BH, $\mathrm{II}$) to being arranged into fields and spacetime (that is what happens, eventually, in $\mathrm{I}$). Now we can formalize that statement, by inverting Eq. \eqref{n1} and getting
\be
\si(N_\I)=\arcsin \sqrt{\frac{N_\I}{N}} \, .
\ee
When $N_I$ is constrained to be an integer $N_I=m$, the $\si(N_\I)=\si_m$ is discretized. Therefore, the evolution parameter is just a way of counting how many modes jumped out.  It cannot be regarded as time, which should emerge, like space, at low energy from $X$ons dynamics. The von Neumann entropy as a function of $\si=\si_m$ is reported in Fig.1.

We could then expect that at each step the number of BH/environment modes was fixed. What is the meaning of fluctuations of $\hat{N}_\I$ and $\hat{N}_\II$ on $|\psi(\si)\ran$? A direct computation shows that
\be
\Delta N_\I(\si) \ = \ \Delta N_\II(\si) 	\ = \ \frac{\sqrt{N} \, \sin (2\si)}{2} \, ,
\ee
where $\Delta N_j$ $\equiv$ $\sqrt{\lan \hat{N}_j^2 \ran_\si-\lan \hat{N}_j \ran^2_\si}$ is the standard deviation of $\hat{N}_j$ on $|\Psi(\si)\ran$.  In agreement with the general results of Ref. \cite{varent}, Fig. 3 clearly shows that this is a measure of the entanglement.  Moreover, for $N=1$, $\lf(\Delta N_j (\si)\ri)^2$ is proportional to the \emph{linear entropy} or \emph{impurity}
\be
\lf(\Delta N_j (\si)\ri)^2\ = \  2 \, S^j_L(\si) \, , \quad j \ = \  \mathrm{I}, \mathrm{II} \, ,
\ee
defined as \cite{NieChu,BJV}
\be
S^j_L \ \equiv \ 1 - \mathrm{Tr} \rho_j^2 \, .
\ee
Note that $\Delta N_j$ can be easily discretized as explained above.
\begin{figure}[htb]
		\includegraphics[width=0.50\textwidth]{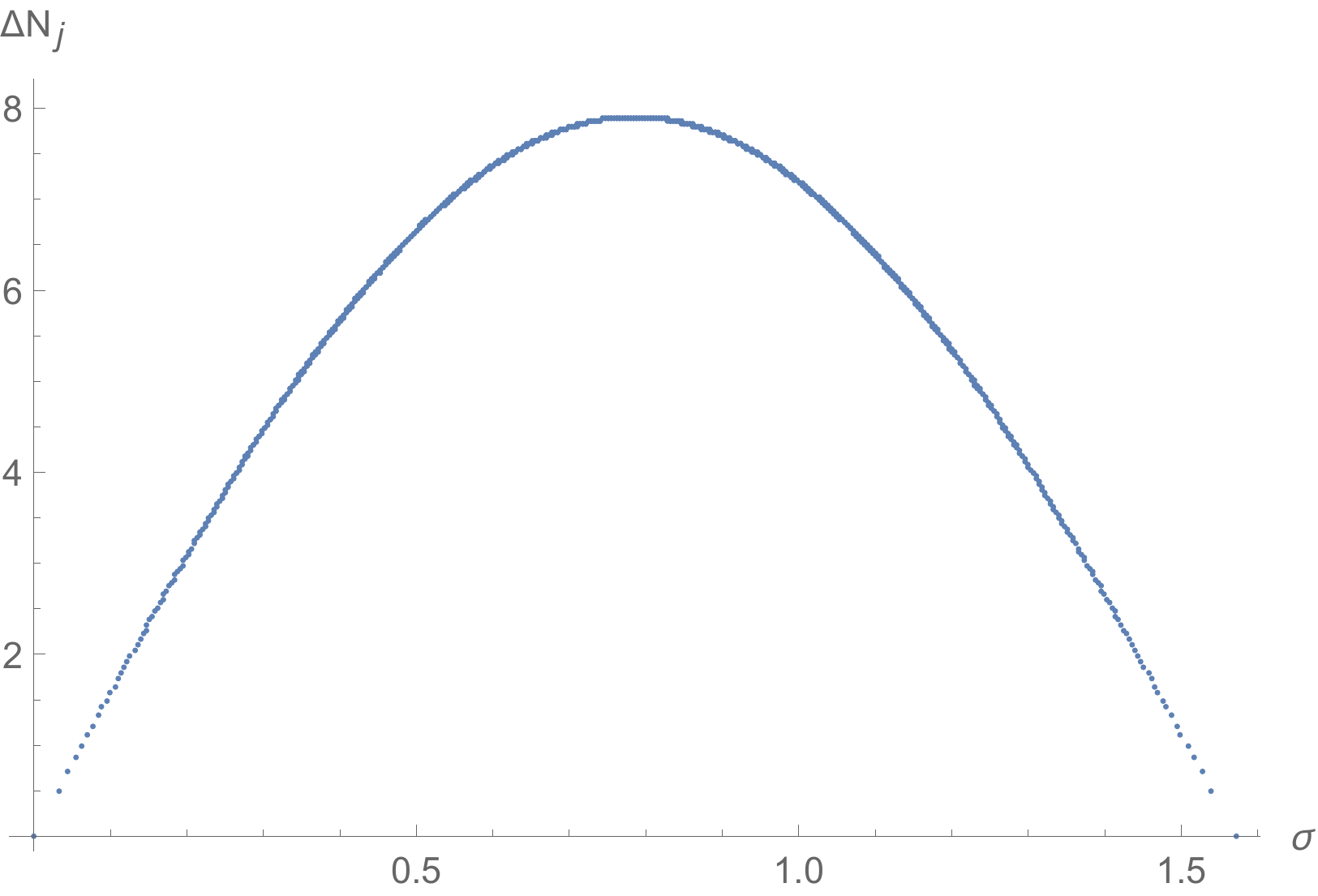}
	\label{fig:fentropy2}
	\caption{$\Delta N_j$ as a function of $\si$, in the case $N=1000$.}
\end{figure}
We finally turn our attention to the generator of the canonical transformations in \eqref{rot1} and \eqref{rot2} (with our choice of parameters)
\bea
a_n(\si) & \equiv & d_n(\si) \ = \ e^{-i \, \si \, G} \, a_n(0) \, e^{i \, \si \, G} \, , \\[2mm]
b_n(\si) & \equiv & c_n(\si) \ = \ e^{-i \, \si \, G} \, b_n(0) \, e^{i \, \si \, G} \, ,
\eea
where one can easily check that
\bea \label{ham2}
G \ = \   i \sum^N_{n=1} \, \lf(a^\dag_n \, b_n \, - \, b^\dag_n \, a_n \ri) \, .
\eea
With the above recalled limitations, the existence of such unitary generator is guaranteed by the Stone--von Neumann theorem and, in the general meaning of \cite{Rovelli:2004tv}, it can be seen to enter the \emph{Wheeler--DeWitt} equation
\be \label{wdw}
H \, |\Psi(\si)\ran \ = \ 0 \, ,
\ee
with $H \equiv i \, \partial_\si-G$. This constrains the kinematical Hilbert space $\mathcal{H}_\I \otimes \mathcal{H}_\II$ to the physical Hilbert space $\mathcal{H}$ as previously extensively commented. Let us remark that for $\si=\si_m$, Eq. (\ref{wdw}) becomes a linear difference (recursion) equation.
\section{Connection with dissipative systems} \label{QuantumDissipation}
In the previous Section we have shown how our toy model possesses a Page-like behavior for entanglement entropy and this can be easily computed by means of the TFD entropy operator. We now ask if this is a mere coincidence or if the connection with TFD can be made more precise.

Let us perform the canonical transformation
\be \label{can}
A_n \ = \ a_n \, , \qquad B_n \ = \ b^\dag_n \, .
\ee
This is not a Special Bogoliubov transformation \cite{BlazRip}. In fact, this transformation can be obtained from
\bea
A_n & = & a_n \, \cos \theta_n \ - \ b^\dag_n \, \sin \theta_n \, , \\[2mm]
B_n & = & b^\dag_n \, \cos \theta_n \ + \ a_n \, \sin \theta_n \, ,
\eea
for $\theta_n=0$. Then it is not connected with the identity. However, we can still define vacua in the new representation
\be
A_n\, |0\ran_{A} \ = \ B_n \, |0\ran_B \ = \ 0 \, .
\ee
One can check that
\bea
|0\ran_A & = & |0\ran_\I \, , \\[2mm]
|0\ran_B & = & |1_1 \, 1_2 \, \ldots \, 1_N \ran_\II \, .
\eea
The second relation follows from the fact that $(b^\dag_n)^2=0$. Therefore, the generator \eqref{ham2} becomes
\bea
G \ = \    i \, \sum^N_{n=1} \, \lf(A^\dag_n \, B^{\dag}_n \, - \, B_n \, A_n \ri) \, .
\eea
This is nothing but the (fermionic version of the) interaction Hamiltonian of quantum dissipative systems, as introduced in Ref. \cite{CeRaVi} (see also \cite{qQuantDiss}). This operator noticeably coincides with the generator of a Special Bogoliubov transformation. Therefore $|\Psi(\si)\ran$ has a TFD-vacuum-like structure \cite{UmeTak,Umezawa}
\bea
|\Psi(\si)\ran & = & \prod^N_{n=1} \, \lf(\cos \si \, + \, \sin \si \, A^\dag_n B^\dag_n\ri) \, |0\ran_A \, \otimes \, |0\ran_B \non \\[2mm]
& = & e^{- \frac{1}{2}  S_A(\si)} \, |I\ran  \, ,
\eea
with $|I\ran \ = \ \exp\lf(\sum^N_{n=1} \, A^\dag_n \, B^\dag_n\ri) |0\ran_A \, \otimes \, |0\ran_B$, and the entropy operators
\bea
S_{A} &=&  -\sum^N_{n=1} \, \lf(A^\dag_n \, A_n \, \ln \sin^2 \si + A_n \, A^\dag_n \, \ln \cos^2 \si\ri) \, , \\[2mm]
S_{B} &=& -\sum^N_{n=1} \, \lf(B^\dag_n \, B_n \, \ln \sin^2 \si + B_n \, B^\dag_n \, \ln \cos^2 \si\ri) \, .
\eea
Therefore, through (\ref{can}), we have now the \textit{usual} entropy operators of TFD \cite{UmeTak,Umezawa,IorLamVit}, to be compared with the \textit{unusual} definitions of \eqref{s2}: $S_\I \ = \ S_A $ and $S_\II \ = \ S_B$.

The physical picture here is that, when the system evolves, a pair of $A$ and $B$ particles is created. The B-modes enter into the BH, annihilating BH modes, while the A-modes form the environment. This mechanism is heuristically the same as the one proposed by Hawking \cite{Hawking}, and lately formalized via the tunneling effect \cite{Tunnel}.

The $A-$ and $B-$modes do not discern explicitly fields and geometric dof.  However, taking the view of Ref. \cite{Acquaviva:2017xqi}, some dof are indeed responsible for the reduction of the BH's horizon area during the evaporation and \emph{annihilators of geometric modes} can be defined. In order to make this idea more precise we can decompose $A_n$ in their geometric ($G$) and field ($F$) parts as follows
\bea \label{d1}
A_n \ = \ \sum_k \, \lf(g_{k,n} \, A^k_{G,n} \otimes  \ide_{F,n}+ f_{k,n} \, \ide_{G,n} \otimes A^k_{F,n}\ri) \, ,
\eea
where $k$ labels the emergent modes. Eqs.\eqref{d1} can be regarded as a dynamical (Haag) map at linear order \cite{Umezawa}.  The full dynamical map -- available once the quantum theory of gravity is specified -- should connect the fundamental dof to the emergent notions of geometry and fields. The coefficients of the map should then lead to the Hawking thermal behavior of the latter, at emergent level. Note that the action of $A_n^\dag$ on $|0\ran_A$, creates both a matter mode and a geometric mode, outside the horizon: the region of spacetime surrounding the BH and available to an external observer increases, because the horizon area decreases. 

Let us remark that, in this picture, a relationship of the form \eqref{d1} makes sense only for $A$-modes (or, equivalently $a$-modes) and not for $B$-modes (or, equivalently $b$-modes). In fact, by definition, the former are the ones which rearrange to form matter and geometry, in contrast with the latter which describe free $X$on. As previously stated, such a distinction is at the basis of our intrinsic notion of interior/exterior of the BH. A more detailed analysis of this issue will be addressed in a forthcoming work.

It is only once we have the notion of area that we could try to fix $N$ in terms of the BH parameters, $\mathcal{M}_0$, $\mathcal{Q}_0$, $\mathcal{J}_0$, and of the Planck length $l_P$. In fact, as remarked in Refs. \cite{Hod:1998vk, maggiore_2008}, a quantum of BH horizon area measures
\be
\De A \ = \ \al \, l_P^2 \, ,
\ee
where $\al$ is a constant to be fixed. Therefore, the BH entropy assumes the form\footnote{This is equal to ${\cal S}_{BH}$, defined in Eq.\eqref{esbh}, only before the evaporation, when ${\cal S}_{BH}={\cal S}_{max}$.}
\be \label{sbhal}
{\cal S}_{BH} \ = \ \frac{\al \, N}{4} \, ,
\ee
where $N$ is the number of Planck cells that, as remarked in the Introduction, would correspond to the number of our quantum levels/slots. 

The value of $\al$ was fixed to $\al=4 \ln 3$, in Ref. \cite{Hod:1998vk}, by means of arguments based on Bohr's correspondence principle, and to $\al=8\pi$ in Ref. \cite{maggiore_2008}, by means of arguments based on the BH quasi-normal modes. In our case, a comparison between \eqref{sbhal} and \eqref{smax} gives
\be
\al \ = \ \ 4 \, \ln 2 \, .
\ee
This value agrees with the condition $\al=4 \ln k$, with $k$ integer, which was proposed in Ref. \cite{Hod:1998vk}, in order to constrain the number of microstates $\Si$ to an integer (see Eq. \eqref{dimh}).  Therefore, our model gives $k = 2$.  A comparison with the standard Bekenstein formula, when $\mathcal{Q}_0 = 0 = \mathcal{J}_0$, gives
\be
N \ = \ \frac{4 \, \pi \, \mathcal{M}^2_0}{l_P^2\, \ln 2} \, ,
\ee
which, of course, is again Eq.\eqref{immirziLQG} when the area is expressed in terms of ${\cal M}_0$ and the Immirzi parameter is fixed to $\ln2 /(\sqrt{3}\pi)$.  In fact, both derivations -- ours and LQG's -- rely on the identification of the entropy with $S_{BH}$.  Clearly, a more detailed analysis, beyond these qualitative arguments, will be possible only once a complete dynamical map will be available.
\section{Conclusions}

We assumed here that the dof of a BH are finite in number and fermionic in nature, and hence obey a Pauli principle. Then, within the approach of second quantization, we naturally obtained that the BH evaporation is a dynamical mechanism producing a maximal entanglement entropy, equal to the initial entropy of the BH. This phenomenon is an instance of the Bekenstein bound, obtained here with arguments that do not assume pre-existing spatiotemporal concepts. Of course, for a full identification with the standard formulae (see, e.g., \cite{Casini_2008,Bousso:2002ju}), one needs to link geometrical concepts, such as elementary Planck cells, to such fundamental dof.

We then showed that entanglement, Bekenstein and environment (thermodynamic) entropies here are all naturally obtained in the same approach, based on an entropy operator whose structure is the one typical of TFD. Through such operator, we have evaluated the von Neumann BH--environment entropy and noticeably obtained a Page-like evolution.

We finally have shown that the latter is a consequence of a duality between our model and a dissipative-like fermionic quantum system, and hence it has a natural TFD-like description.

Many directions for further research need be thoroughly explored, the most important being a reliable dynamical map from the fundamental modes to the emergent fields/spacetime structures.  This would pave the road for the investigation of important aspects of the emergent picture, such as the Hawking thermality of the field sector.  Moreover, a geometric interpretation of the results presented here would allow to investigate the discreteness of the spectrum of quasinormal modes, which is known to be related to the quantization of the BH horizon's area (see Ref.\cite{maggiore_2008}). Anyway, even in the absence of a full dynamical map, we believe that our simple, although nontrivial, considerations are necessary to fully take into account the fascinating and far-reaching consequences of the Bekenstein bound.
\begin{acknowledgments}
The authors would like to thank G. Vitiello, P. Luke\v{s}, D. Lanteri and L. Buoninfante for enlightening discussions. A. I. and L. S. are partially supported by UNiversity CEntre (UNCE) of Charles University in Prague (Czech Republic) Grant No. UNCE/SCI/013.
\end{acknowledgments}

\appendix
\section{Equivalence of TFD entropy and von Neumann entropy} \label{vntfd}

In this Appendix we explicitly show the equivalence of the TFD entropy of Eq.\eqref{s12} with the von Neumann entropy of Eq.\eqref{svn}.  We first present the computation in the simplest cases of Eqs.\eqref{1psi} and \eqref{2psi} and then we shall focus on the computation of the von Neumann entropy.  Let us note that the expectation value of the TFD entropy operators (see Eq.\eqref{s12}) immediately follows from Eq.\eqref{numav}.
\begin{itemize}
\item
In the case $N=1$, the density matrix reads
\bea \non
&& \rho(\si) \ = \ \cos^2 \si \, |0\ran_\I |1\ran_\II \, {}_\II \!\lan 1 | {}_\I \! \lan 0| +\sin^2 \si \, |1\ran_\I |0\ran_\II \, {}_\II \! \lan 0 | {}_\I \! \lan 1| \\[2mm]
&& +\ \frac{\sin (2\si)}{2} \lf( |0\ran_\I |1\ran_\II \, {}_\II \!\lan 0 | {}_\I  \!\lan 1|+|1\ran_\I |0\ran_\II \, {}_\II \!\lan 1 | {}_\I \! \lan 0|\ri) \, ,
\eea
where we omitted tensor product symbols. The reduced density matrices have the following form:
\bea
 \rho_\I(\si) & = & \cos^2 \si \, |0\ran_\I \, {}_\I \! \lan 0| +\sin^2 \si \, |1\ran_\I \, {}_\I \! \lan 1| \, , \\[2mm]
 \rho_\II(\si) & = & \cos^2 \si \, |1\ran_\II \, {}_\II \! \lan 1| +\sin^2 \si \, |0\ran_\II \, {}_\II \! \lan 0| \, .
\eea
Eq.\eqref{svn} follows immediately from these expressions.
\item
In the case $N=2$, we directly report the reduced density matrices:
\bea \non
&& \rho_\I(\si) \ = \ \cos^4 \si \, |0 0\ran_\I \, {}_\I \! \lan 0 0| +\sin^4 \si \, |1 1\ran_\I \, {}_\I \! \lan 1 1| \\[2mm]
&& +\ \frac{\sin^2 2\si}{4} \lf(  |0 1\ran_\I \, {}_\I \! \lan 0 1|+ |1 0\ran_\I \, {}_\I \! \lan 1 0|\ri) \, , \\[2mm]
&& \rho_\II(\si) \ = \ \cos^4 \si \, |1 1\ran_\II \, {}_\II \! \lan 1 1| +\sin^4 \si \, |0 0\ran_\II \, {}_\II \! \lan 0 0| \non \\[2mm]
&& +\ \frac{\sin^2 (2\si)}{4} \lf(  |0 1\ran_\II \, {}_\II \! \lan 0 1|+ |1 0\ran_\II \, {}_\II \! \lan 1 0|\ri) \, .
\eea
It follows that
\bea
\mathcal{S}_\I(\si) & = & -\sin^4 \si \, \ln \sin^4 \si-\cos^4 \si \, \ln \cos^4 \si \non \\[2mm]
&-& 2 \, \sin^2 \si \, \cos^2 \si \, \ln \lf(\sin^2 \si \, \cos^2 \si\ri)\non \\[2mm]
& = & \mathcal{S}_\II(\si) \, .
\eea
By using the relations $\ln (a b) \ = \ \ln a + \ln b$ and $\cos^2 \si+\sin^2 \si =1$, we get
\bea
\mathcal{S}_\I(\si) & = & -2\lf(\sin^2 \si \, \ln \sin^2 \si+\cos^2 \si \, \ln \cos^2 \si\ri) \non \\[2mm]
& = & \mathcal{S}_\II(\si) \, ,
\eea
which is equal to Eq.\eqref{svn} for $N=2$.
\end{itemize}
One could repeat similar computations for all $N$. However it is simpler to use the correspondence of our model with a TFD/dissipative system via Eq.\eqref{can}. As already known in TFD, the ``thermal vacuum'' can be rewritten in the form
\be
|\Psi(\si)\ran \ = \ \sum_{n=0,1} \, \sqrt{w_n(\si)} \, |n\ran_A |n\ran_B \, ,
\ee
where $|n_A\ran_A$, $|n_B\ran$ are eigenstates of the number operators
\be
\hat{N}_{A} \ = \ \sum^N_{n=1} \, A^\dag_n \, A_n \, , \qquad \hat{N}_{B} \ = \ \sum^N_{n=1} \, B^\dag_n \, B_n  \, .
\ee
Moreover, the coefficients $w_n$ are given by
\be
w_n(\si) \ = \ \prod^N_{j=1} C_j^{2} (\si) \, . \label{wn}
\ee
and $C_j$ were firstly introduced in Eq.\eqref{alexp}.  The density matrix thus reads
\be
\rho(\si) \ = \ \sum_{n=0,1} \, w_n(\si) \, |n\ran_A |n\ran_B \, {}_B \! \lan n| {}_A \! \lan n| \, ,
\ee
from which the reduced density matrices are easily derived:
\bea
\rho_A(\si) \ = \ \sum_{n=0,1} \, w_n(\si) \, |n\ran_A \, {}_A \! \lan n| \\[2mm]
\rho_B(\si) \ = \ \sum_{n=0,1} \, w_n(\si) \, |n\ran_B \, {}_B \! \lan n| \, .
\eea
The von Neumann entropy reads \cite{UmeTak,IorLamVit,Dias:2019ezx}
\be
\mathcal{S}_A(\si) \ = \ \mathcal{S}_B(\si) \ = \  - \, \sum_{n=0,1} w_n(\si) \, \ln w_n(\si) \, .
\ee
Finally, Eq.\eqref{svn} follows from substituting the explicit form of $w_n$ (cf. Eq.\eqref{wn}) into the last expression.

\end{document}